# Jupiter's North Equatorial Belt and Jet: III. The 'great northern upheaval' in 2012

John H. Rogers & Gianluigi Adamoli

## Summary

In Paper II we described the extreme changes in Jupiter's North Equatorial Belt (NEB) which took place in 2011-12: quiescence, narrowing, and fading of the belt, along with acceleration of the prograde NEBs jet to super-fast speed. Here we describe how this anomalous state was terminated in 2012, in a rapid and vigorous disturbance known as a NEB Revival – the first in living memory. At the same time, the North Temperate Belt (NTB) had entered a similar preparatory state, which was terminated by a NTB Revival initiated by a typical outbreak on the super-fast NTBs jet. The two concurrent Revivals appeared to constitute a 'great northern upheaval' that extended from the equator to ~38°N, which may have been the first such event ever recorded. We compare this dual event with the more typical 'Global Upheavals', which consist of near-simultaneous Revivals of the NTB and the South Equatorial Belt (SEB), along with equatorial coloration, and we discuss the current understanding of Global Upheavals.

## Introduction

Conventions and abbreviations are as in Paper I [ref.1] and Paper II [ref.2]. Drift rates in longitude are given as DL1 (deg/day or deg/month [30 days] in System I) or DL2 (deg/month in System II) or $u_3$ (m/s in System III). Latitudes are planetographic. South is up in all illustrations.

The sources for this account are our BAA/JUPOS analyses for 2011/12 and 2012/13, as already presented in our reports on-line [ref.3]; these include more extensive descriptions and illustrations. A summary was given in ref.4. The observers for 2012/13 are listed on our web site [ref.3g].

## The NTBs jet

The very fast jet on the NEB south edge (NEBs) was described in Papers I and II. The jet on the NTB south edge (NTBs) is at least as fast, and was undergoing changes in apparent speed which would come to a head in the NTB Revival at the same time as the NEB Revival in spring, 2012. Historically this jet was only observable when there were distinct outbreaks of spots on it [ref.5], but with spacecraft data, and with very-hi-res amateur imaging since 2003, it has been possible to measure the wind speed even when there was no outbreak occurring. This jet has two very different states (Figure 1).

In the 'fast' state, which held from 1991 to 2005 and in several earlier outbreaks of dark spots, it carries vortices at DL1 ~ -60 deg/mth ($u$ ~ 125 m/s, called the North Temperate Current C or NTC-C), and the peak wind speed of the jet at cloud-top level is only ~10 m/s faster [refs. 6-8]. In the 'super-fast' state, which held from 1970 to 1990 and again from 2007 onwards, it undergoes spectacular outbreaks initiated by one or more brilliant white plumes at DL1 ~

-160 deg/mth (*u* ~ 170 m/s, called the North Temperate Current D or NTC-D) [refs.5, 9-15]. In such an outbreak chaotic dark spots appear following the super-fast plumes, leading to the developmemt of a vivid orange NTB(S), and usually to the revival of the whole NTB. These outbreaks occurred every 5 years from 1970 to 1990 (assuming one during solar conjunction in 1985); the only previous comparable outbreak had been recorded in 1880; and they resumed in 2007. These outbreaks have been described in detail for 1975 [refs.9,10], 1990 [refs.11,12], and 2007 [refs.13, 14]. Each proceeded in the same way, as summarised elsewhere [refs. 5, 8, 16].

Usually such outbreaks occur when the NTB has faded, and lead to revival of the dark belt, so they are sometimes called 'NTB Revivals'. This is not always accurate: for example the NTB was not faint prior to the 1975 event, whereas it sometimes revived without a jet outbreak in previous decades. Nevertheless, the term has suited the phenomenon in recent decades, and is also appropriate as it resonates with the names of comparably grand Revivals of the SEB and NEB. As we shall discuss below, these three types of Revival have much more in common than their diverse appearances had suggested.

A. Sanchez-Lavega's group have modelled the atmospheric dynamics of the NTBs jet outbreaks, and find that they require the jet below cloud-top level to have a permanent super-fast speed of at least 170 m/s [refs.14-16]. During 'fast' (NTC-C) eras, this deep jet is necessary to sustain the vortices at cloud-top level, but these impose a strict limit of ~135 m/s on the speed at that level. When the vortices disappear, the super-fast speed can spread up to cloud-top level, reducing the stratification of the atmosphere so that a convective disturbance deep down in the water clouds can propagate up to the surface, in the form of the spectacular super-fast plumes. After the outbreak, the cloud-top speed immediately collapses back to the NTC-C range, but our evidence is that during 'super-fast' (NTC-D) eras it recovers within a few years, in time for another outbreak five years later (Figure 1).

This scenario is supported by our observations in recent years. As the NTB whitened and the vortices disappeared in 2003-2004, tiny streaks allowed us to track the NTBs jet itself, which gradually accelerated [ref.7; Fig.1]. In early 2007, observations by New Horizons and the Hubble Space Telescope showed that the jet had reached super-fast speeds, and in 2007 March, a classic super-fast outbreak began. The speed collapsed back to the 'fast' range as that outbreak developed [refs.13, 14, 17]. But we then found the speed increasing again in 2010 (Fig.1), and based on that measurement by G. Hahn, and on the historical 5-year periodicity, we predicted another super-fast outbreak in 2011-12 [ref.18].

## The great northern upheaval

### *I. The start in spring, 2012*

At the start of 2012, both the NEB and NTB were faded [Paper II]. The general calm was dramatically broken in 2012 March and April, just before solar conjunction, with violent outbreaks in the both belts [Refs. 3b,3c].

The NEB outbreak began on 2012 March 8 with a small bright spot in NEBs and small dark blue-grey spot on its S edge; they remained associated, with DL1 = +1.8 deg/day, until March 14 (Figure 2). By then the dark blue-grey spot had become a conspicuous, very dark projection, while the bright spot had expanded into a long rift which moved more slowly. On March 12, a second bright spot appeared p. the first (at ~9.5°N); a dark blue-grey spot appeared on its S edge on March 14 and repeated the behaviour of the first one (DL1 = +2.0 deg/day). The outbreak was first reported by Wayne Jaeschke on March 18, with the two dark spots accompanied by a bright rift. The two dark spots became intensely dark and large, and remained so in the last image on April 6; they were at 7.6 (±0.6) °N. However by this time there was still no rapid expansion of the disturbance and no revival of the NEBn; we suspected that these might follow.

We had also been expecting a NTBs jet outbreak in 2012, as explained above [ref.18]. Indeed the outbreak began in mid-April, as seen in the last few images before solar conjunction (Figure 3). It was discovered by Manos Kardasis in his images on 2012 April 19, as a bright spot on the NTBs with a very dark spot f. it. By comparison with the 2007 outbreak, it had probably started around April 12, although it had not been visible in a lo-res image on that date. On April 21, Gianluigi Adamoli also recorded these features, consistent with the expected super-fast drifts (DL1 ~ -5 deg/day). The very last images probably showed the f. end of the outbreak on April 26. All the data indicated that a typical super-fast outbreak had begun, but it was impossible to determine accurate speeds or to resolve the features.

## *II. The climax in summer, 2012*

Early images in the next apparition (beginning with Kardasis again on 2012 June 4) showed that both outbreaks had developed impressively. The best images from June are in Figure 4. At first glance the North Tropical and Temperate domains appeared chaotic; but closer study confirmed that classical Revivals of the NEB and NTB were indeed occurring. Although most of the developments had been hidden during solar conjunction, it was still possible to identify key features of the Revivals. Figure 5 shows maps from historical examples in which the same features can be identified: an NEB Revival in 1893, and a NTB Revival in 2007. Maps of the northern hemisphere throughout July are aligned in Figure 6.

In June and July there were many dark formations on the NEBs edge including striking dark spots; the NEB was very broad, though pale and chaotic; and reddish or ochre colour covered much of the NEB, NTropZ, and NTB(S). The whole region was very complex. There was also an ochre band in central EZ, probably associated with the NEB Revival; conversely, the NNTB latitudes were almost white. Thus the unique occurrence of great disturbances on the NEB and NTB could be called a 'great northern upheaval', affecting all latitudes from 0 to ~38°N.

Events proceeded rapidly. By the end of July, sectors of the belts and the intervening NTropZ which were still light in June had filled in with intense turbulence and reddish (ochre or orange) colour, producing one vast brown-and-ochre belt from the NEBs to the NTB. Large-scale outbreaks had stopped, and the disturbance had evolved from larger to smaller scales, so the whole region was filled with small streaks, which formed a remarkable herring-bone pattern across the whole region in hi-res images, directly revealing the wind gradients.

By late August the disturbed regions were already settling down, with a very broad NEB and an orange NTB(S); both belts had indeed fully revived. The following detailed account is adapted from Refs.3d,e,f.

*Equatorial Band:*

There was a notable orange-brown Equatorial Band at most longitudes in June-July, with some dark grey streaks in it. This was the most distinct equatorial coloration event since 1989-91, and was presumably associated with the NEB Revival in some way. It spanned latitudes ~2°N to S. No distinct source for the reddish colour could be identified. In late June one stretch of EB (in the sector where NEBs projections were not yet well formed) was quite dark brown, and as they did form, in July, the EB cleared from this sector. For over a month the ochre EB had a fairly distinct p. end with positive DL1, but it was unclear whether this represented the colour being cleared away by white plume clouds emanating from the NEBs, or just represented the usual drift rate on the equator. By August, the ochre EB only existed around about half the circumference, and was getting paler and shorter, although fainter patches of similar colour existed at other longitudes.

*Northern EZ & Southern NEB:*
In June, the appearance was bizarre, but entirely consistent with an on-going NEB Revival (Figsures 4 & 5)  The most remarkable sector was from L2 ~ 50-130, which was very pale, but hi-res images on June 19-20 (Figure 4) showed it to be filled with complex light spots and streaks.  This may have been an intensely convective 'rifted' region comprising the source of the outbreak. It soon filled in, so at the start of July the whole NEB was light brown and filled with turbulent streaks, but with no distinct rifts.

Typical dark grey or bluish-grey NEBs dark formations (NEDFs) were already seen in June around half the circumference, moving approximately with System I (L1).  Some were very dark.  These may have been a series generated as the NEBs jet flowed past the rifted region, as we saw commencing in March. More developed in the remaining half during July. Relative to System II longitude (L2), the leading NEDFs would have spread all around the planet by July, while the trailing ones were still being formed. The array of NEDFs extended round the whole belt by the end of July.  They were at ~8°N, and the NEBs edge itself was unusually far north at 9°N.

These NEDFs were quite well-formed and stable, and some were very large.  Whereas the first NEBs dark spots in March-April had highly positive DL1 (Figure 2b), in July and August these spots had normal drifts: mean DL1 = +0.17 deg/day, range -0.23 to +0.5 deg/day. By early August there was a stable array of 13 of them around the planet. The main activity was between the spots numbered 14 and 1 and 2, which were repeatedly splitting to produce new, small, very dark, short-lived spots that retrograded with DL1 ~ +0.5 to +1.6 deg/day.  This behaviour is typical alongside intense disturbances within the NEB, which tend to generate and partially entrain dark spots on or near the NEBs edge.

Within the NEB in July there was ubiquitous small-scale turbulence, but no distinct rifts nor bright spots (except a pair of bright streaks that were moving with DL2 = -3.6 deg/day, from late July onwards). Bright spots would reappear in late August (e.g. one in Figure 6), perhaps representing the resumption of normal activity.

*Northern NEB and N.Tropical Zone:*
In June there were large-scale contrasts between different sectors, and several very dark grey spots at roughly fixed L2 in the NTropZ (see below) (Figure 4). These are very unusual for the NTropZ, but similar spots had been seen during previous NEB broadenings or revivals (e.g. one in 1893: Figure 5.). But by early July, no obvious large features persisted (Figure 6).  There were many dark patches scattered all across these latitudes (18-21°N) but they were rapidly changing.

Later in July, the main dark grey features were innumerable streaks lying obliquely across the northern NTropZ (20-22°N), indicating the wind speed gradient. (Similar streakiness was also seen after the NTBs outbreak in 2007, so it may be due to either or both outbreaks.)  Meanwhile the NTropZ also increasingly filled in with dusky ochre shading, mixed with the herring-bone pattern of grey streaks.  In the less streaky sectors of NTropZ, hi-res images confirmed that there was definitely diffuse colour, which was not just due to low resolution (e.g. Figure 7).

After a broadening or revival of the NEB, an array of cyclonic brown ovals ('barges') and anticyclonic white ovals (AWOs) normally forms [ref.5].  This had already begun in August, with several surprising aspects (Figures 8 & 9):

(i) At least one barge and one AWO seem to have survived from the previous year.  The AWO was the well-known old White Spot Z (WSZ) (Figure 8); but it was a dark grey spot in June, and remained largely shrouded in grey streaks into August.  (WSZ had also been dark grey for a time during the 2007 NTBs outbreak [ref. R13b].)

(ii) Some of the AWOs were actually transformations of the dark spots seen in June! In June there were four very dark grey spots and one fainter one, roughly stationary at L2 = 109 (spot d4), 193 (d7, the least conspicious), 219 (d8), 274 (d9), and 359 (d5-d6, a long bar). Of these five dark spots, 3 or 4 were replaced by white spots, and detailed inspection of images shows what happened (Figures 8 & 9). Spots d7, d8 and d9 were each replaced by a small white spot in early July, apparently displacing the dark grey material into streaks around it. However, these white spots were again masked by dark grey streaks later in July, only to reappear in August, drifting faster (White spots A and Z). Spot d4 disappeared around June 20, but we suspect that it also reappeared as an AWO (White spot B) on the same track.

It is also notable that changes in appearance of these three AWOs occurred within days of each other, even though they were far apart.

This remarkable transformation of anticyclonic dark grey spots into white spots in these latitudes has not been documented before, to our knowledge (apart from the temporary changes to WSZ in 2007), but it fits in with what we know of these features. Dark grey or brown spots sometimes appear in the NTropZ in the early stages of a NEB expansion event or Revival, and are evidently anticyclonic rings; for instance, we documented several during the last broadening event in 2009. So no change in circulation is needed to convert them into AWOs. We may speculate that the very dramatic and rapid transformation in this year, along with the unusually early appearance of barges, could be related to the greater intensity of this full-blown Revival in contrast to more modest broadening events.

(iii) A new array of barges as well as AWOs was already forming in August, only 5 months after the start of the Revival (Figures 8&9), in addition to one brown barge (at L2 ~ 125) which probably survived from 2011. Like the white spots, the barges began very small, still buffetted by the surrounding turbulence, but nevertheless were visible even in modest images. More small AWOs and barges appeared in the autumn. But by 2013 Feb., almost all the AWOs had merged or disappeared; the last two large ones (WSA and WSZ) merged in Feb. [refs.20 ,21]. At least 3 pairs of barges also merged [ref. 21a]. By 2013 March, there were just six barges and two AWOs including WSZ.

*NTB(S):*

By June the NTB(S) had already revived as a dark orange-brown belt (23 to 26°N) around most of the circumference, as is typical after a super-fast outbreak. In one sector ~100° long, it had not yet revived but comprised dark and bright spots – probably the tail end of the super-fast outbreak, though we could not measure any drift rates. (See Figure 5, map in 2007 May, at a similar time after the start of the outbreak, and similar to 2012 June.)

The orange-brown NTB(S) was still very prominent in July, although there was still a short gap in it. Within this gap were indistinct grey streaks which could not be tracked, but we were able to obtain approximate drift rates for several features flanking the gap. These were:

**(a)** The p. end of the dark orange-brown NTB f. the gap: DL1 = -69 deg/month;

**(b)** A reddish blob (anticyclonic oval?) on the NTBs: DL1 ~ -30 deg/month in late July, decelerating to DL1 ~ +23 deg/mth in early August, when it had shifted to just south of the NTBs;

**(c)** The f. end of a small white strip on NTBs : DL1 ~ -44 deg/month. Similar speeds are normally recorded for many dark spots in super-fast outbreaks, but were probably missed this time during solar conjunction.

Several findings confirmed that a typical super-fast outbreak had occurred, even though the initial stage was only seen fuzzily for a few days in April, and no distinct spots remained after solar conjunction with super-fast speed (which would be DL1 ~ -5 deg/day). The following features in the later stages were typical, just as in 2007 [Ref. 13b].

(i) The orange NTB(S);

(ii) The turbulence across the NTropZ;

(iii) Super-fast speed estimated for a small reddish condensation within the belt, which moved in L1 by ~-3 deg in 10 hours on July 15;

(ii) Speeds of DL1 ~-1 to -2 deg/day observed for several features (see above]). These speeds are typical of the secondary phase of a super-fast outbreak.

*NTB(N):*
A narrow dark grey NTB(N) existed all round the planet, at least from July onwards (29ºN to ~31-34ºN), but was discontinuous, with many streaks and wiggles. It was separated from the orange NTB(S) by a light strip which seemed to show intense small-scale rifting in hi-res images (Figure 7). The JUPOS team tracked many NTB(N) features at all longitudes, and they found typical N. Temperate Current A drifts, DL2 = +16 to +29 deg/month. The speeds were much the same as in 2011, but the actual features were different. (In subsequent years, this belt became much more wavy and also developed into a N. Temperate Disturbance, as it did in the years after 2007. [refs.22, 23])

*NNTB:*
The NNTB was essentially invisible in 2012 July-August, being replaced by a broad white zone from NTZ to NNTZ (Figure 10). This rare occurrence was no doubt another consequence of the great upheaval, in two respects. First, there seems to be a general tendency for increased darkening on one domain to be compensated by brightening in an adjacent domain [ref.5 p.248]. Second, super-fast NTBs outbreaks generally suppress activity in the NNTBs jet-stream (see discussion below.)
NNTBs jet-stream activity resumed in 2013 Jan. An oblique turbulent streak had already been present in the whitened NNTB latitudes in autumn 2012, and in late Jan., 2013, it started to emit tiny dark spots south-preceding it in the NNTBs jet-stream – and also a few larger, very dark spots north-following it in the NNTZ [ref. 21b].

*Zonal wind profiles:*
Given the huge changes in apparent speed of the NEBs and NTBs jets during these upheavals, it is of interest to know whether the the zonal wind profile across the NEB changed. During the NEB recession and fading in 2011/12, a well-determined zonal drift profile (Figure 11a) was normal in the NEBn-NTropZ. (It appeared anomalously far south in the mid-NEB, but this may be because it was determined for small dark spots, whereas profiles across these latitudes are normally determined for bright spots in 'rifts'.) Determining drift rates during the Revivals was difficult, given the low resolution of the early images and the complexity of the region, but our tracking of visibly identified features from June to August showed clear gradients of speed across the NEB and NTropZ which agreed closely with the zonal wind profile from spacecraft (Figure 11b). There could be some distortion of the NTBs jet peak, which we also found to be chaotic after the 2007 outbreak. Also, Grischa Hahn (personal communication) derived a zonal wind profile by map correlation analysis in WinJUPOS, using a pair of images on 2012 July 15, and although the results showed some scatter, they were consistent with an unchanged zonal wind profile from the NEBs to the NTropZ; as was a profile from Hubble images in September (Figure 11c).

*Latitudes after the Revivals:*
By 2012 Dec., the NEB and NTB had settled into fairly normal appearance with a narrow bright white NTropZ between them. The NEB was still disturbed, with barges and rifts, and streaky dark material extending to 20ºN (**Table 1, below**), but not uniformly, and long sectors of the northern part were lightening again in early 2013. The NTB consisted of well-defined components now in contact, the NTB(S) still orange and the NTB(N) now dark grey.

## ***III. Multispectral imaging***

Images in the ultraviolet (UV) and the infrared methane absorption waveband (889 nm: $CH_4$) give further information about the changes in the belt patterns, especially the new orange bands. UV images are very sensitive to reddish colours, which appear UV-dark. Methane images show high clouds and hazes as bright, because they are above most of the methane absorption in the atmosphere; it was expected that orange bands would be methane-bright. Analysis of these images was reported in [ref.19], which is summarised here, along with latitude measurements (**Table 1**).

In methane images, surprisingly, the new orange EB was quite dark, though not uniformly so: the darkest parts coincided with blue-grey streaks along its edges, as though these streaks were very methane-dark while the orange haze itself was methane-transparent. In UV images, as expected, the new orange EB was notably dark, and even the sector which was not strongly coloured visually was fairly dark in UV.

The NTB was very faint in methane images from late 2007 to 2011, except for a narrow, streaky, visibly bluish North Tropical Band in 2010 and 2011 (23ºN). It seemed that a reddish methane-bright haze, persisting since the 2007 NTB Revival, obscured the usual dark belt. With the NTB Revival in 2012, the major part of the NTB again became methane-dark as usual (25-32ºN; Table 1). However, the new orange NTB(S) (23-26ºN) overlapped the methane-dark NTBs edge (25ºN); i.e. the southern part of the NTB(S) was methane-bright, but the northern part was not. This was consistent with the same stage in 2007, showing that the new reddish NTB(S) in its early stages is not necessarily methane-bright. Along its northern edge, either the haze is so thin as to be methane-transparent, or it lies deeper than expected.

Overall, our review from 2007 to 2012 [ref.19] showed that there is not a strong correlation between the visible colour and methane brightness. Even if a high-altitude reddish haze is often present – as in the NTB from mid-2007 to 2009 – it is sometimes too thin or too low to appear methane-bright, and it is sometimes accompanied by clearing of the deeper clouds which adds a darker, brown or even grey tone to the visible colour. These events are complex weather patterns that can affect multiple cloud layers in various ways.

**Table 1.** Latitudes of belt edges in UV, visible, and methane bands.

| | UV+B | | | | RGB | | | | CH4 | | | | RGB (2013) | |
|---|---|---|---|---|---|---|---|---|---|---|---|---|---|---|
| | *Mean* | SD | N | | *Mean* | SD | N | | *Mean* | SD | N | | *Mean* | SD |
| SEB(N)n | **-7,6** | 0,4 | 8 | | **-7,8** | 0,3 | 7 | | **-7,6** | 0,8 | 3 | | | |
| EBs | **-2,6** | 0,5 | 5 | | **-2,1** | 0,6 | 4 | | | | | | | |
| EBn | **2,1** | 0,7 | 5 | | **1,6** | 0,1 | 2 | | | | | | | |
| NEBs | **9,0** | 0,5 | 7 | | **7,6** | 0,1 | 2 | @ | **11,2** | 1,2 | 6 | | **8,6** | 0,5 |
| | | | | | **9,3** | 0,2 | 5 | | | | | | **16** | 0,5 |
| NEBn | | | | | | | | | **20,9** | 1,0 | 7 | | **19,9** | 0,2 |
| NTB(S)s | **23,0** | 0,6 | 7 | | **23,2** | 0,5 | 7 | | **25,1** | 0,6 | 8 | | **23,1** | 0,2 |
| NTB(S)n | **26,2** | 0,4 | 8 | | **26,1** | 0,3 | 6 | | | | | | **26,3** | 0,4 |
| NTB(N)s | **29,3** | 0,4 | 4 | | **29,2** | 0,4 | 7 | | | | | | **26,3** | 0,4 |
| NTB(N)n | **31,5** | 0,2 | 4 | @ | **31,8** | 0,9 | 7 | | **31,9** | 0,5 | 5 | | **30,6** | 0,3 |
| | **32,9** | 0,3 | 4 | | | | | | | | | | | |

First three data sets: Table adapted from [ref.19]. Zenographic latitude measurements made by Gianluigi Adamoli on 8 multispectral image sets between 2012 July 11 and August 10. @Values in boxes appeared to be bimodal.
Fourth data set: Measured from 6 RGB maps made by Marco Vedovato, 2013 Jan.-March.

---

## Discussion (1): Effects of NEB and NTB outbreaks on neighbouring zones

*Coloration:*
Reddish coloration is a typical (though not inevitable) consequence of all these major disturbances – SEB Revivals, NEB Revivals, and NTBs super-fast outbreaks – and it is often not restricted to the belt which revived, often spreading into adjacent zones [Ref. 5 p.55]. So it was no surprise that reddish colour spread not only over the NEB and NTB(S), but also over the intervening NTropZ, and around part of the EZ. This was the first time we had hi-res images of such a widespread coloration event, and they confirmed that diffuse colour across the NTropZ was real (e.g. Fig.7), not an artefact of low resolution.

*Suppression of NNTBs jet-stream activity:*
The absence of NNTBs jet-stream spots was notable because they also disappeared immediately after the 2007 NTBs super-fast outbreak, suggesting that this may be a regular connection. Indeed, an historical survey confirms that NNTBs jet-stream spots were also absent for a full year after the NTBs super-fast outbreaks of 1970, 1975, and 1980, and virtually absent after the suspected outbreak of 1985 *[see list below]*. The only exception to this rule was in 1990.

This connection adds to our understanding of global upheavals. As originally defined (see below), a global upheaval typically included a NNTBs jet-stream outbreak starting with or soon after the NTBs jet-stream outbreak (see below). However, the hi-res observations of recent decades have shown nearly continuous NNTBs jet-stream spot activity, although the number of spots varies, so 'outbreaks' of these spots are not as well-defined as previously thought. Possibly the real connection is that NTBs outbreaks (at least of the super-fast type) normally suppress NNTBs jet-stream activity, and what has often been observed is the *re*-appearance of normal activity a year or so later.

**Table 2. Suppression of NNTBs jetstream activity by NTBs outbreaks**

The dates of onset of the NTBs super-fast outbreaks, and the appearance of NNTBs jet-stream spots thereafter, were as follows. [from Ref.3f]

1970 August: No NNTBs jet-stream spots were seen thereafter, nor in 1971, even though there had been plenty up to 1970 July. New NNTBs spots 1972 July.

1975 Sep.: None in 1975 nor 1976. New NNTBs spots 1978 Nov.

1980 May: None in 1980/81. New NNTBs spots by 1984 May.

1985 (suspected NTBs outbreak during solar conjunction early in the year): Just one spot in 1985, and 4 tiny ones in 1986, recorded mainly from the Pic du Midi. New NNTBs spots by 1988 August.

1990 Feb.: The exception. A substantial outbreak of NNTBs jet-stream spots started in 1990 Jan. and continued at least until March, and was still continuing (after solar conjunction) from Dec. onwards. (This was also the only NTBs super-fast outbreak which did not produce substantial reddish colour.)

**Discussion (2): Historical perspective and implications for Global Upheavals**

*Timing of the NEB and NTB outbreaks*

The coincidence of the NEB Revival and the NTBs outbreak (NTB Revival), which started only a month apart, is both satisfying and puzzling. It is satisfying because both occurred in line with previously established periodicities of 3-5 years for NEB expansion events and 5 years for NTBs super-fast outbreaks. But it is puzzling because no previous instances were known when both occurred together. Perhaps the two events coincided merely by chance – or perhaps there was a physical connection between them, but apparently unique to this year.

The historical record of NTBs outbreaks may be incomplete, as the super-fast spots last only a few months, and the smaller spots require high resolution to track. The orange colour of the revived belt may be a more reliable indicator of super-fast outbreaks, and it was not reported between about 1890 and 1960, except in three apparitions [ref.5 p.101]. One of these was 1927-28, and in ***Appendix 2*** we discuss whether this might have followed a dual outbreak in 1926, at the time of the last NEB Revival. In ***Appendix 1*** we summarise more recent events which resembled the 2012 'great northern upheaval', in 2016-17.

*Updated view of Global Upheavals*

More typically, NTBs outbreaks tend to coincide with SEB Revivals, and with EZ coloration, constituting a 'Global Upheaval'. This is a set of large-scale outbreaks of various types in different domains on Jupiter, originally described as a 'zenological disturbance' by Wacker (1975) [ref.24]; the concept was refined and renamed Global Upheaval by Rogers [refs.5,10]. The latter description included, as the most important elements, a SEB Fade/Revival cycle, coloration in the EZ, and – a year or two later -- NTBs and NNTBs jet-stream outbreaks [**Table 3**].

**Table 3. Typical Pattern of a Global Upheaval**

| | |
|---|---|
| T = -1 to -2 yr | GRS rotation period begins to increase |
| T = -1 yr | SEB(S) is very faint, GRS prominent; **± STropD** |
| | EZ becomes ochre with white spots in N |
| T = 0 | SEB Revival, --> GRS fades; ± STBn jet spots |
| | [*new*: **NTBs (NTC-D) jet outbreak**] |
| T = +1 to +2 yr | (SEB becomes reddish) |
| | [*old*: NTBs (NTC-C) & NNTBs jet outbreaks] |
| | [*new*: **NNTBs jet spots resume**] |
| | (NTropZ becomes yellowish) |
| | EZ quietens and becomes more bluish |
| T = +3 to +4 yr | NTC-B appears in NTB. |

The table is taken from ref.5 p.250, with new amendments in **bold**.
**'± STropD'** means that a S. Tropical Disturbance may be present in some instances.
NTC = North Temperate Current (NTC-B, -C, -D are different speed ranges).

The reality of these upheavals has been reinforced by events since Wacker's [ref. 24] proposal. The triple coincidence of SEB, EZ and NTBs events has recurred in 1975, 1990, and 2007; these account for all the SEB Revivals in this period except for one in 1993 and one in 2010, and all the EZ colorations except for one in 1978-80 and a weak one in 2001-02. (The years 1978-1980 might also qualify as a Global Upheaval as the EZ coloration was accompanied by vigorous activity within the SEB and on some jets, including a NTBs outbreak, but no SEB Revival.)

During these 40 years, with these very distinct, well-separated events, and hi-res observations from spacecraft and, latterly, from amateur imaging, we have come to understand these events better. All the southern hemisphere events in the Global Upheaval schedule are now recognised as attributes of the SEB Fade/Revival cycle, and S. Tropical Disturbances, which have occurred in a few such cycles, may also be included among these [refs.26, 27]. So is the EZ coloration (although it can also appear in association with NEB cycles, as in 2012). Now we can identify changes in the NNTB (especially the suppression of NNTBs jet activity) as consequences of the NTB cycle. So the Global Upheaval now appears to be, essentially, a coincidence of SEB and NTB cycles, along with EZ coloration.

Moreover, we can now see that the SEB Revival outbreak and the NTBs jet outbreak typically occur only months apart, much closer in time than was previously thought. So this sharpens the definition of the Global Upheaval. We now see that most clear examples involved a NTC-D outbreak, which often occurred within just months before or after the SEB Revival outbreak (1970, 1975, 1990, 2007). In contrast, two 'classical' global upheavals (1928-29, 1938-39) now look unusual in that the NTBs outbreaks displayed NTC-C only (there was no orange colour, so probably no NTC-D outbreak), and started more than a year after the SEB Revival outbreak [ref.5].

*Similarities between Revivals of the SEB, NEB and NTB*

Revivals of the SEB, NEB and NTB not only have a tendency to occur close in time, they also involve surprisingly similar physical processes, according to recent understanding -- in spite of the apparent differences between them. The SEB and NTB outbreaks both start with a convective upsurge which can rise up from a deep level (probably the hypothetical water cloud layer) only when special conditions prevail in the cloud-top layer; and a similar process may be involved in NEB expansion events. The cycles in these three belts have recently been compared by Fletcher [ref.25]. Moreover, the Great White Spot storms on Saturn seem to be similar. [refs.30-34]

In more detail, the Fade-Revival cycles of the SEB, NTB, and NEB show the following similarities.

(i) Each cycle sometimes recurs with periodicity of 3-5 years, over time-spans which can be as long as 20-30 years, separated by intervals in which such outbreaks are rare or absent. [ref.5]

(ii) In a preparatory phase or 'Fade', the vertical structure of belt changes (in terms of cloud cover, temperature, and/or zonal speed at cloud tops), with visible whitening of some latitudes as white cloud cover develops over the usually dark belt. Similarities between the Fade stage in the SEB and NEB were discussed in Paper II.

-- For the SEB, the Fade is preceded by an end to convective activity in the belt [ref.26]. Also, changes in the vertical temperature profile have been monitored [ref.28], showing a cooling at the start of the Fade; we do not know if this also occurs in the NEB and NTB.
-- For the NEB, the 2012 example suggests that disappearance of both rifts and NEDFs may have initiated the Fade, while also allowing the permanent deep super-fast jet to extend up the the surface.
-- For the NTB, disappearance of slow NTBs vortices allows a permanent deep super-fast jet to extend up the the surface, which is a prerequisite for the convective outbreak to start [refs.7, 14-16].

(iii) A vigorous convective plume, rising from the water layer to above the cloud-tops, initiates the outbreak which will then spread around the belt and lead to its revival. In each case, there can be more than one such event, at remote longitudes. But curiously, the plume(s) arise in a different dynamical location for each type of outbreak:
-- In the SEB, it arises in a stationary point (near-zero drift) in the zonal wind profile (ZWP), at least sometimes being in a small cyclonic oval locally at rest; this then becomes a stationary source from which successive plumes arise and prograde, forming a series of convective cells [refs.27,29].
-- In the NTB, it arises in the peak of the prograde jet, with the fastest speed that is ever seen at Jupiter's cloud-tops [refs. 8,14,16].
-- In the NEB, it arises within the strongly sheared belt, as a white spot which becomes a 'rift'. For NEB Revivals, we have no information other than that the 2012 Revival began with a southerly and fast-moving white spot. On the other hand, NEEs begin with northerly and slow-moving white spots, which then elongate and proliferate (although it is not clear whether they actually initiate the broadening process) [Paper II].
-- Saturn's Great White Storm of 1990 arose in the equatorial prograde jet;
-- Saturn's Great White Storm of 2010 arose in the peak of a retrograde jet [refs.32-34].

(iv) Subsequent changes show individual characteristics for each belt, as the whitish cloud cover is removed and the belt returns to normal; the process involves 'rifts' (streaks of convective white clouds), dark streaks, and circulations or waves.
--Within the SEB and NEB, there is large-scale rifting which evolves to smaller scales; in the NTB, intense but small-scale rifting as the NTB(N) develops [ref.13b; & Fig.7; & Juno images in 2017]. There may then be a quiescent interval before normal rifts reappear in each belt. In the NTB, a more distinct rifted region often develops when the Revival is complete [refs. 22, 23].
--On the low-latitude edge of the belt, in many cases, relatively slow-moving 'spots' are induced on the super-fast jet. On the SEBn, this was not a feature of the 2007 and 2010 Revivals so has not been well characterised. On the NEBs, the slow-moving 'spots' are NEDFs as reported in Paper II and herein; on the NTBs they are vortices [refs.6,15,16].
--On the high-latitude edge of the belt, various conspicuous phenomena may represent surplus energy being taken up into prominent circulations or waves, viz: SEBs jet spots; NEBn broadening then barges and AWOs; NTBn waves then very dark spots and (with the aforementioned rifting) sometimes a N. Temperate Disturbance [refs.22, 23].

(v) Reddening. This was an well attested sequel of historic SEB and NEB events [ref.5], but has not been properly assessed in the modern era in the absence of objective colour measurements from images. For the NTB, the appearance of a new orange NTB(S) is an obvious outcome of a NTC-D outbreak.

Thus, Global Upheavals are now a well-established phenomenon defined by near-simultaneous occurrence of SEB and NTB cycles with equatorial coloration, while the 'great northern upheaval' of 2012 was the only known coincidence of NEB and NTB cycles (but with possible parallels in 1926-27 and 2016-17; see Appendices). At present we have no theory to explain these apparent linkages between cycles in different domains. But evidence for very deep circulation patterns within Jupiter, beginning to emerge from the Juno orbiter as this article is written, may perhaps enable such a theory to be developed in the not-too-distant future.

---

## *Appendix 1:* Another Great Northern Upheaval in 2016-17

Since the 'great northern upheaval' of 2012, the saga has continued with further cycles of outbreaks on both the NEB and NTB. Both of them continued the established periodic sequences, but with anomalies in timing that led to near-simultaneous disturbances again in 2016-17. In early 2015, the author issued a 'Three-year weather forecast for Jupiter' [ref.35], suggesting that a new NEB expansion event (NEE) would start imminently, and a new NTBs jet outbreak would start in late 2016 or early 2017.

For the NEB, the prediction was based on the 3-year periodicity and the appearance in late 2014 of slow-moving rifts, as described in Paper I of this series. In the event, there was some NEBn activity, but it was only during solar conjunction in late 2015 that a NEE really began; then in spring, 2016, having extended 140° around the planet, it stalled and regressed [Refs. 36-38]. It seemed that this NEE had failed. But in autumn, 2016, rifts again appeared in the NEB, including some slow-moving northerly ones, and proliferated extensively during the winter, leading to a rapid NEE affecting all longitudes in 2017.

For the NTB, the prediction was based on the 5-year periodicity, the faintness of the belt, and the speed of the jet in 2014-15. In fact, a classic super-fast jet outbreak did occur, but only 4.5 years after the previous one; it probably started at solar conjunction in mid-September, 2016, and was discovered in mid-October at the time of the Juno spacecraft's second perijove [refs. 16, 39-41]. There were four super-fast plumes, the last of which disappeared at the end of October, and the outbreak proceeded in the same manner as previous ones reviewed herein. By November the NTropZ was intensely disturbed, and this turbulence may well have induced the rifting in the NEB that initiated the NEE. From December onwards there was revival of a vividly orange NTB(S) and of a turbulent NTB(N) that then became dark grey.

So at the climax of these disturbances in spring 2017, there was again intense disturbance all across the NEB, NTropZ and NTB, which looked very much like the great northern upheaval in 2012 [ref.42]. However, whereas the 2012 events in the NEB and NTB were both clearly prepared well in advance and were initiated in keeping with their independent periodicities, in 2016-17 the NEE was unexpected (coming soon after an abortive one) and may have been triggered by turbulence from the NTBs outbreak.

As these events occurred just as NASA's Juno orbiter mission was getting under way, they elicited strong interest and collaboration from professional scientists, and were also monitored by the camera and microwave radiometer on Juno itself.

---

## *Appendix 2:* Was there a Great Northern Upheaval in 1926-1927?

Has there ever before been a ‘great northern upheaval’ as in 2012? There are no definite examples, as NEB Revivals and NTB Revivals (of the NTC-D outbreak type) generally did not occur during the same decades [ref. 5]. The only occasion when observations suggest there might have been one was in 1926-1927. These years were described in the BAA Memoirs, as summarised in [ref. 5], and can be supplemented by colour drawings by the German observer Walther Löbering (Fig.12). His are the only coloured portrayals that I have been able to locate from these years, and they confirm the BAA colour descriptions; they show meticulous observation, even though they were mostly not finished to publication quality. They have not previously been published, and scans have been kindly provided by Frau Ulrike Toll and Hans-Jörg Mettig. The only previous publication was [Ref.43], but did not include these drawings from the 1920s.

The NEB was unusually narrow in 1925, and became very disturbed in mid-1926, which I interpret as a classic NEB Revival, though it continued to progress for more than a year. By summer 1927, the NEB was quite broad and very red, but still broadening, with intense rift activity and arrays of white and dark spots on NEBn – the classic outcome of a NEB Revival.

Meanwhile the NTB was dark grey in 1925 and ‘cold grey’ or bluish grey in 1926. In 1926 June-Sep. it carried at least one small dark NTBs jet spot, but this was typical of the NTC-C activity; super-fast bright spots were never observed. Nevertheless, in late 1927, many longitudes of the NTB were clearly ‘ruddy’, as confirmed in Löbering’s drawings (it would return to dark grey in 1928). This was one of only 3 apparitions in a 70-year span when the belt appeared brown or reddish, so it suggests the possibility of an unobserved NTC-D outbreak during solar conjunction between autumn 1926 and mid-1927. Likewise the NTropZ, which had been bright (though creamy-coloured) in 1926, was dull and ruddy at some longitudes in late 1927 and more generally in late 1928. This reddish shading of the narrowing NTropZ could have been spillover from the slowly broadening NEB, but it particularly resembles the appearance during the simultaneous NEB and NTB Revivals of 2012 and 2017.

---

## Figures (miniature copies)

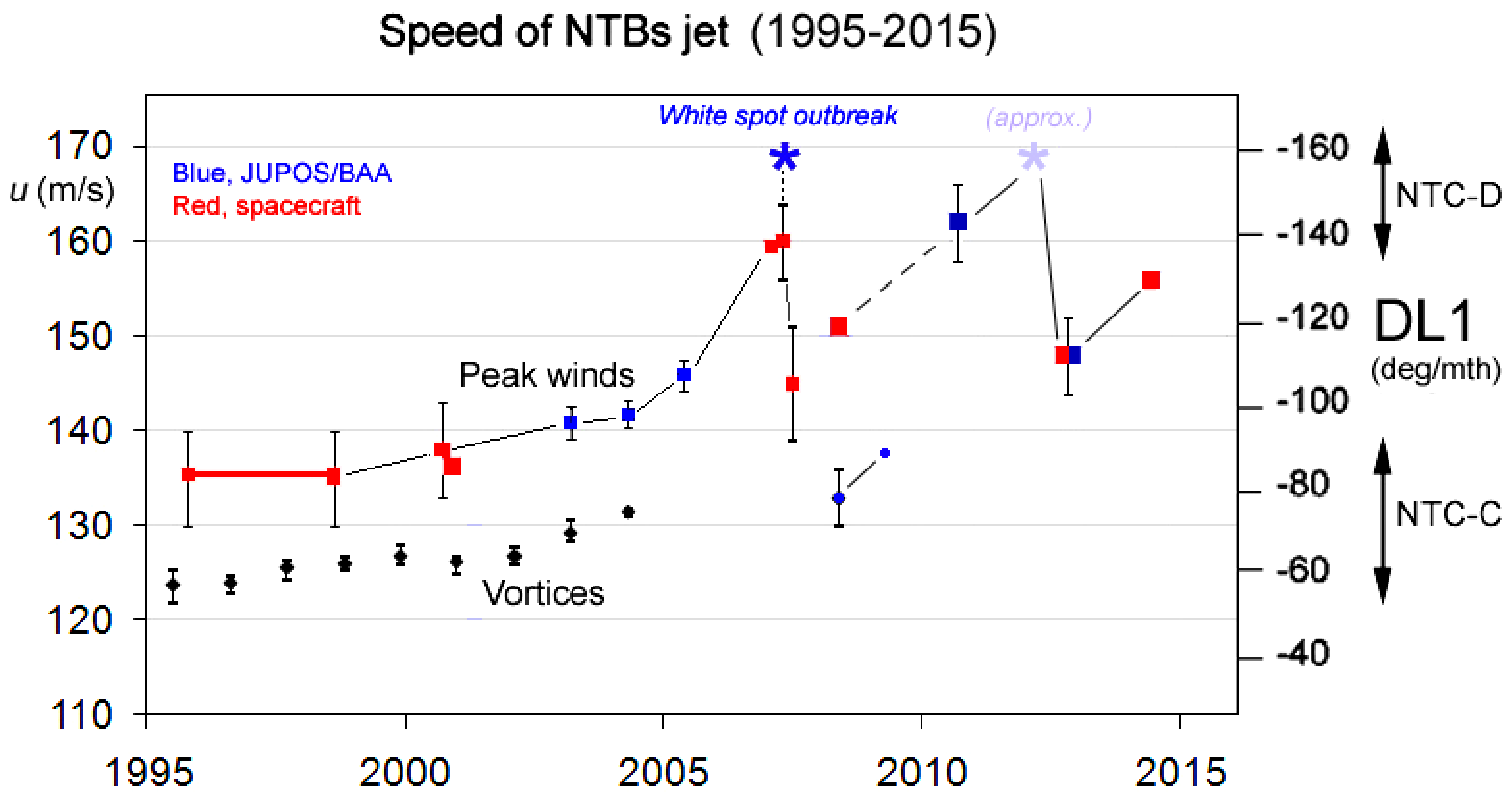

**Figure 1.** Speed of the NTBs jet since 1995. This is from Ref.35, extending the chart in Ref.8, q.v. for sources of data. (Also see profiles of the jet peak from Hubble data, in Ref.17.)

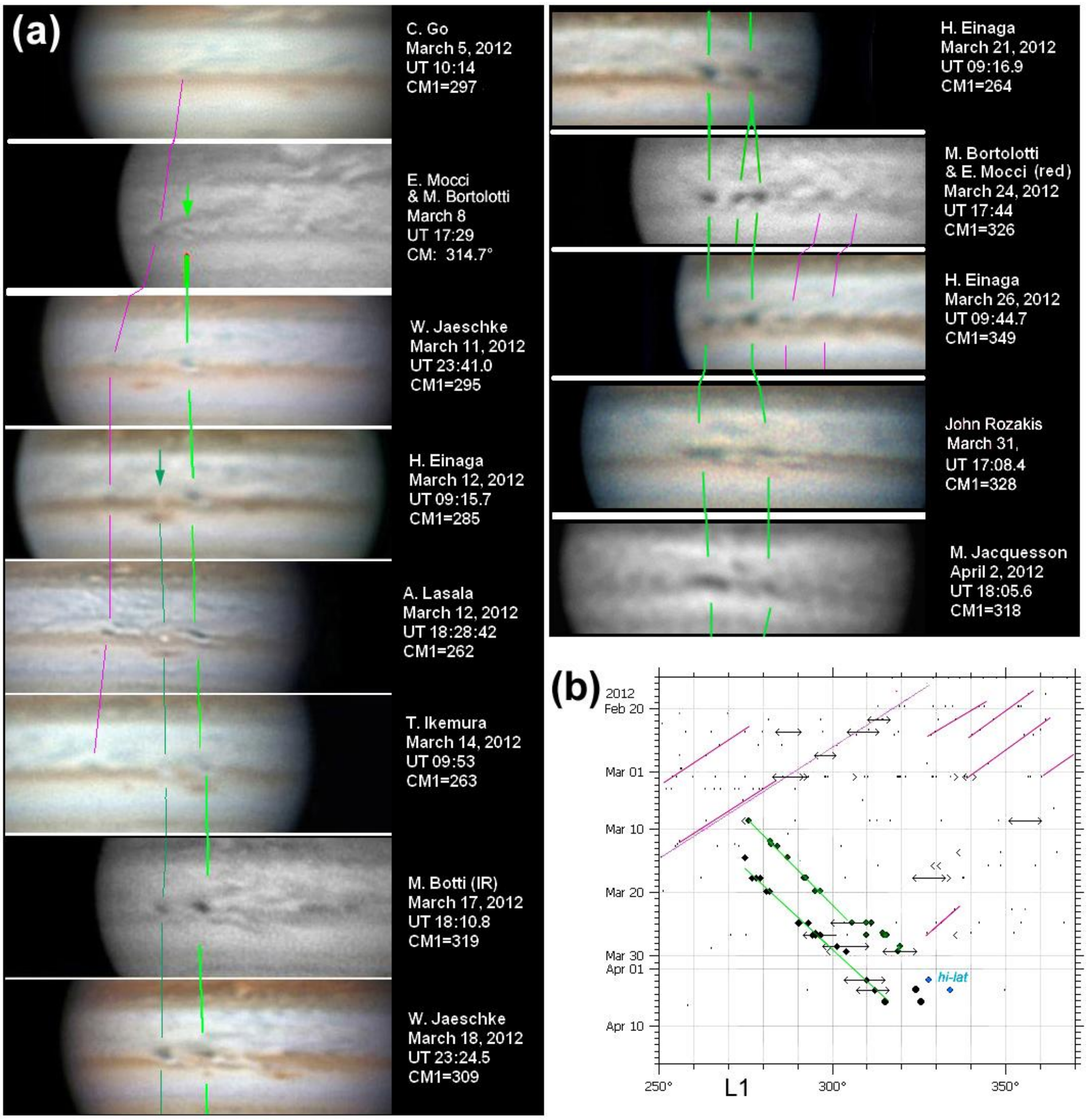

**Figure 2.** (a) Images of the start of the NEB outbreak in 2012 March, selected from a larger compilation by the authors plus Y. Iga in Ref.3b. South is up in all images. Purple lines mark some of the many super-fast NEBs projections with DL1 ~ -2.8 deg/day that preceded the outbreak. Green lines mark the new bright spots and then the very dark NEBs spots which developed alongside them, with DL1 ~ +1.8 to +2.0 deg/day. (b) JUPOS chart of L1 vs time for these NEBs dark spots.

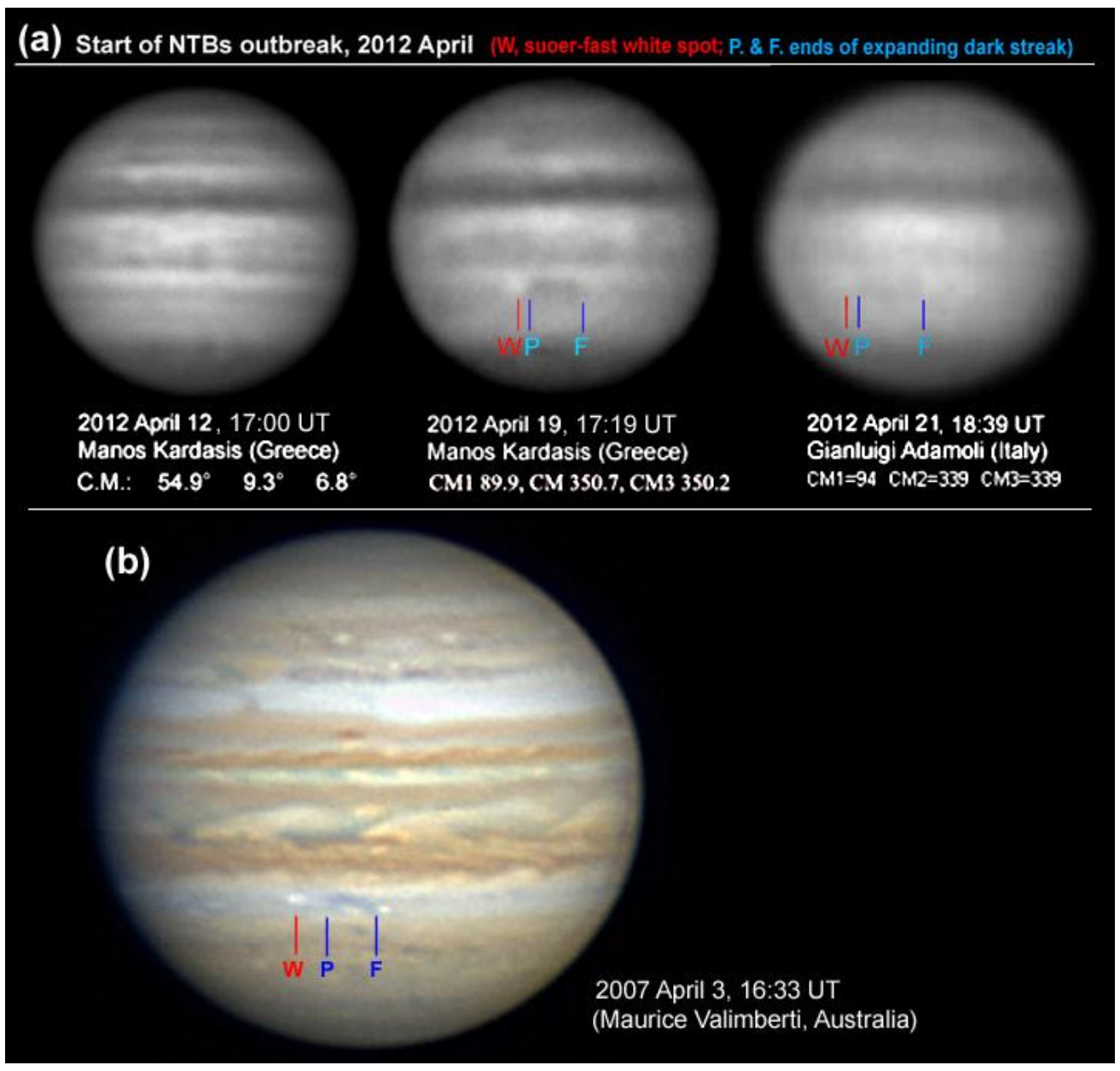

**Figure 3.** (a) Images before and during the start of the NTBs outbreak in 2012 April, from Ref.3c. (The first two are in near-infrared.)
(b) Image on 2007 April 3, similar to 2012 April 19 but at higher resolution. This shows one of two plumes ('W') that had appeared on 2007 March 27.

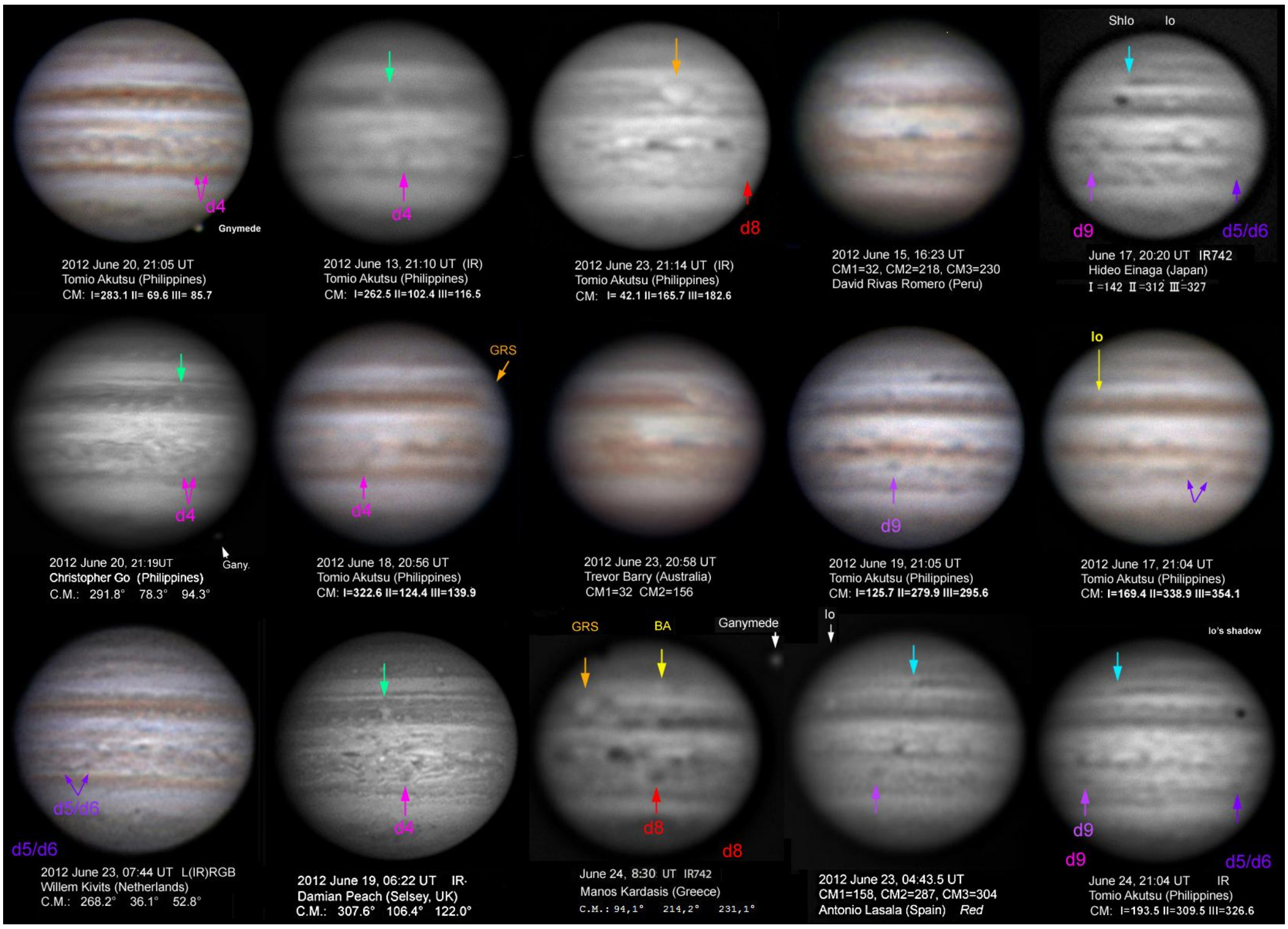

**Figure 4**. Colour and IR images from 2012 June, arranged in order of L2. Major features are marked, each with a different coloured arrow. Adapted from a larger set in Ref.3d.

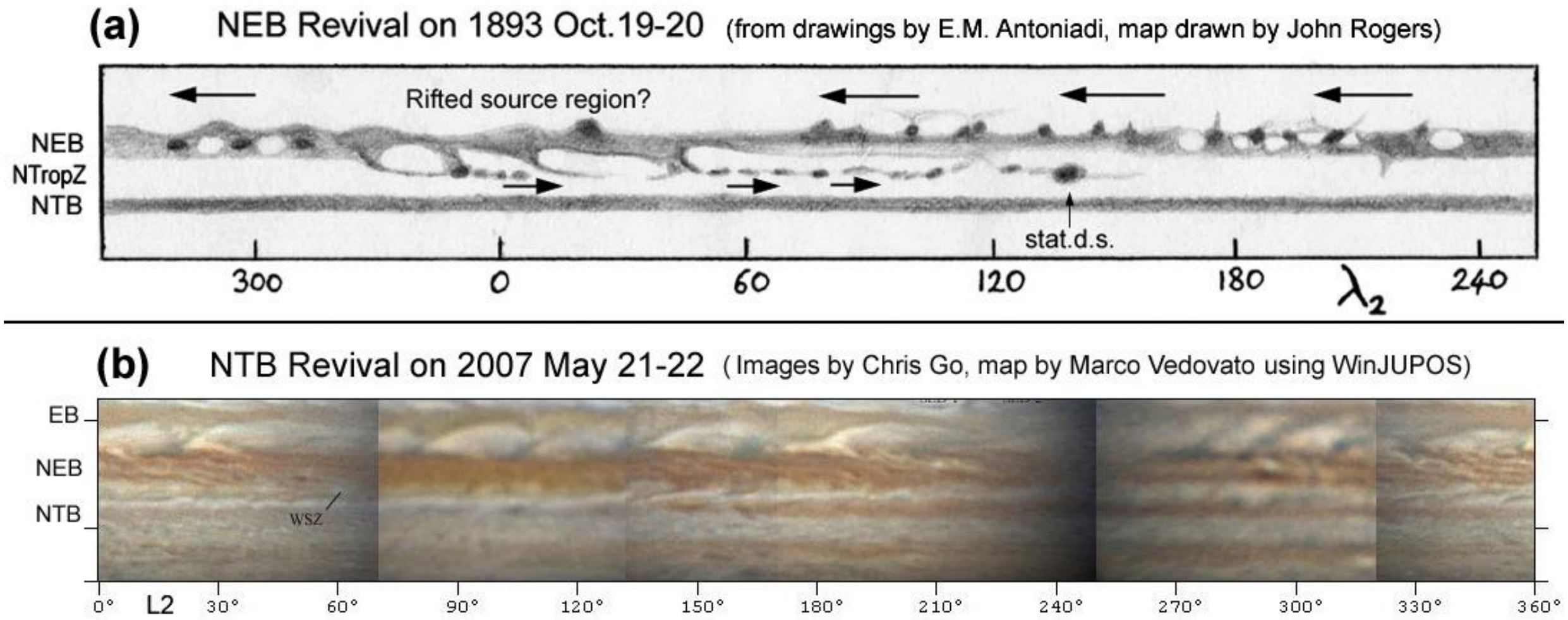

**Figure 5.** Maps of historical outbreaks, comparable to the situation in 2012 June.
(a) NEB Revival on 1893 Oct.19-20 [adapted from Ref.5]. The map suggests that the source of the outbreak was a rifted region around and preceding L2 ~ 45, with dark spots on NEBs and NEBn geenrated from it. Arrows indicate suggested drifts. Note a possible stationary dark spot in NTropZ.
(b) NTB Revival on 2007 May 21-22 [previously unpublished]. At L2 ~ 80-140, grey streaks remain from the outbreak. NTB has revived as a massive orange-brown belt at L2 > 145.

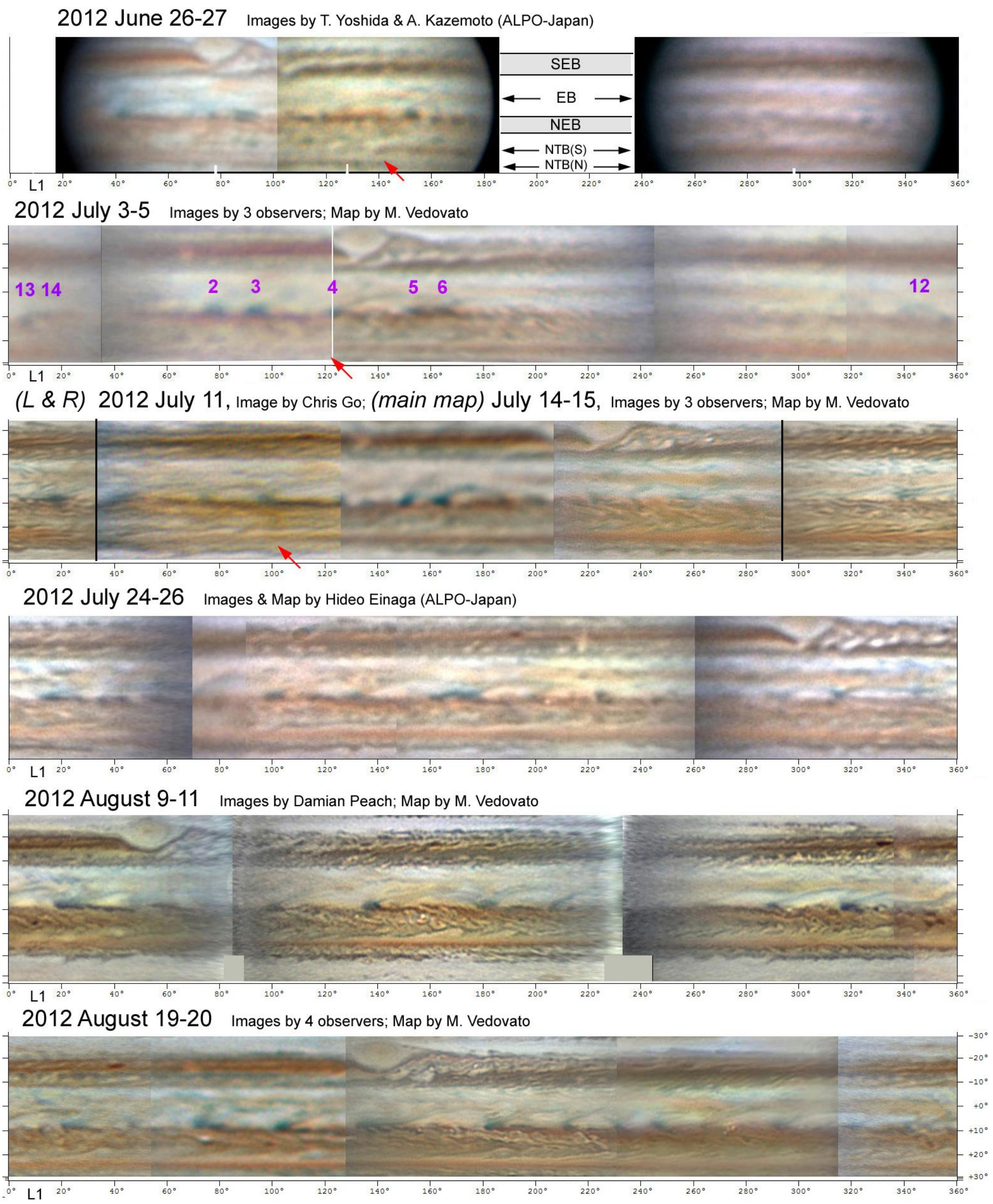

**Figure 6.** Maps in 2012 July, aligned in L1. Prominent NEBs dark formations are numbered in the second map. Red arrow marks the tapered p. end of the dark orange NTB(S). Adapted from a larger set in Ref.3f.

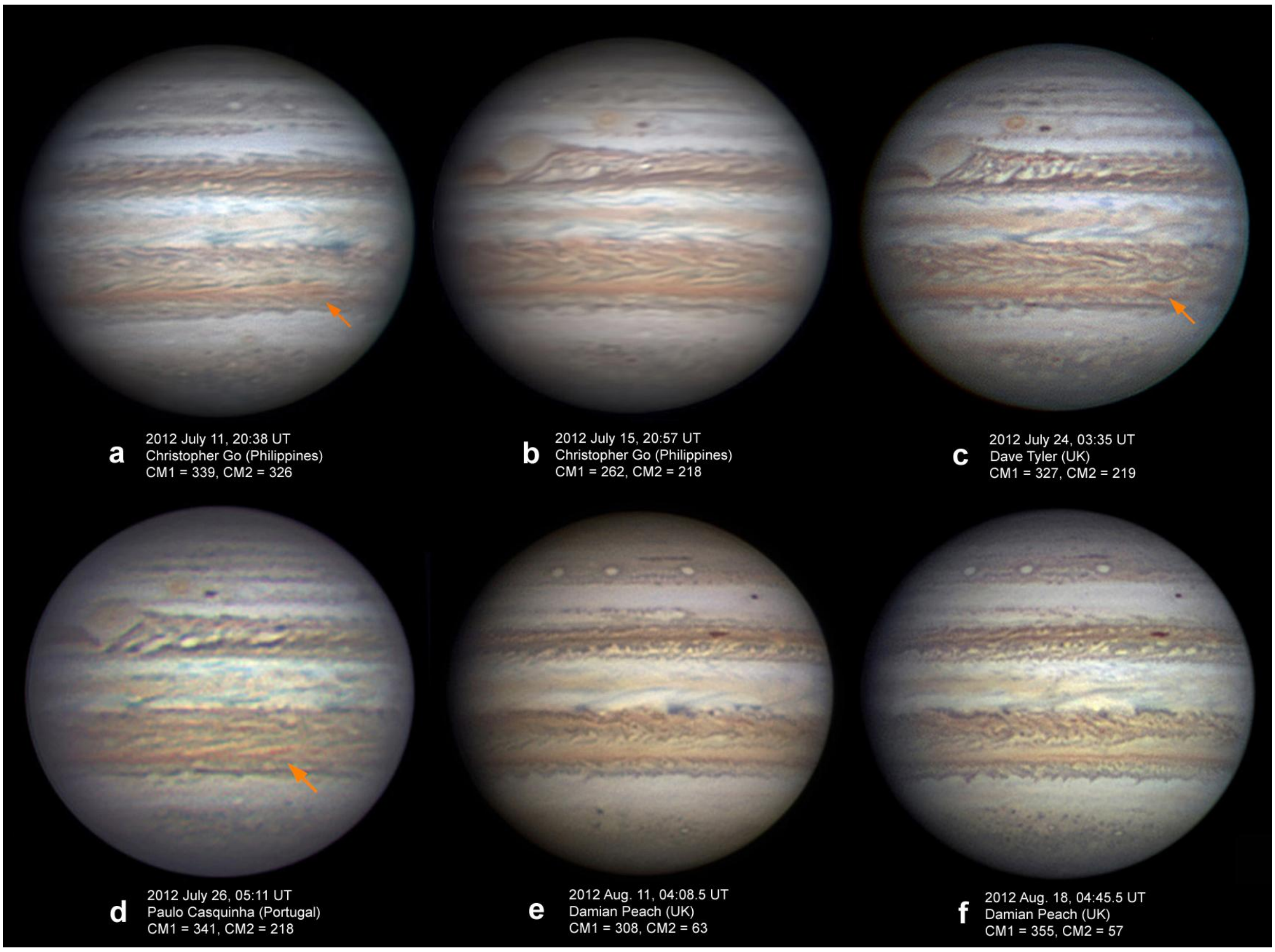

**Figure 7.** Some of the best images from 2012 July-August (also see Ref.4). In (a, c & d), orange arrow marks a dark reddish blob on NTBs with DL1 ~ -1.0 deg/day, in a darkening sector of NTropZ. Images (b, c & d) show white spots w7 and w8, which probably merged to form WSA, and (on f. side of (c & d) the emerging WSZ; see **Fig.8** for key. Image (e) shows WSB on f. side.

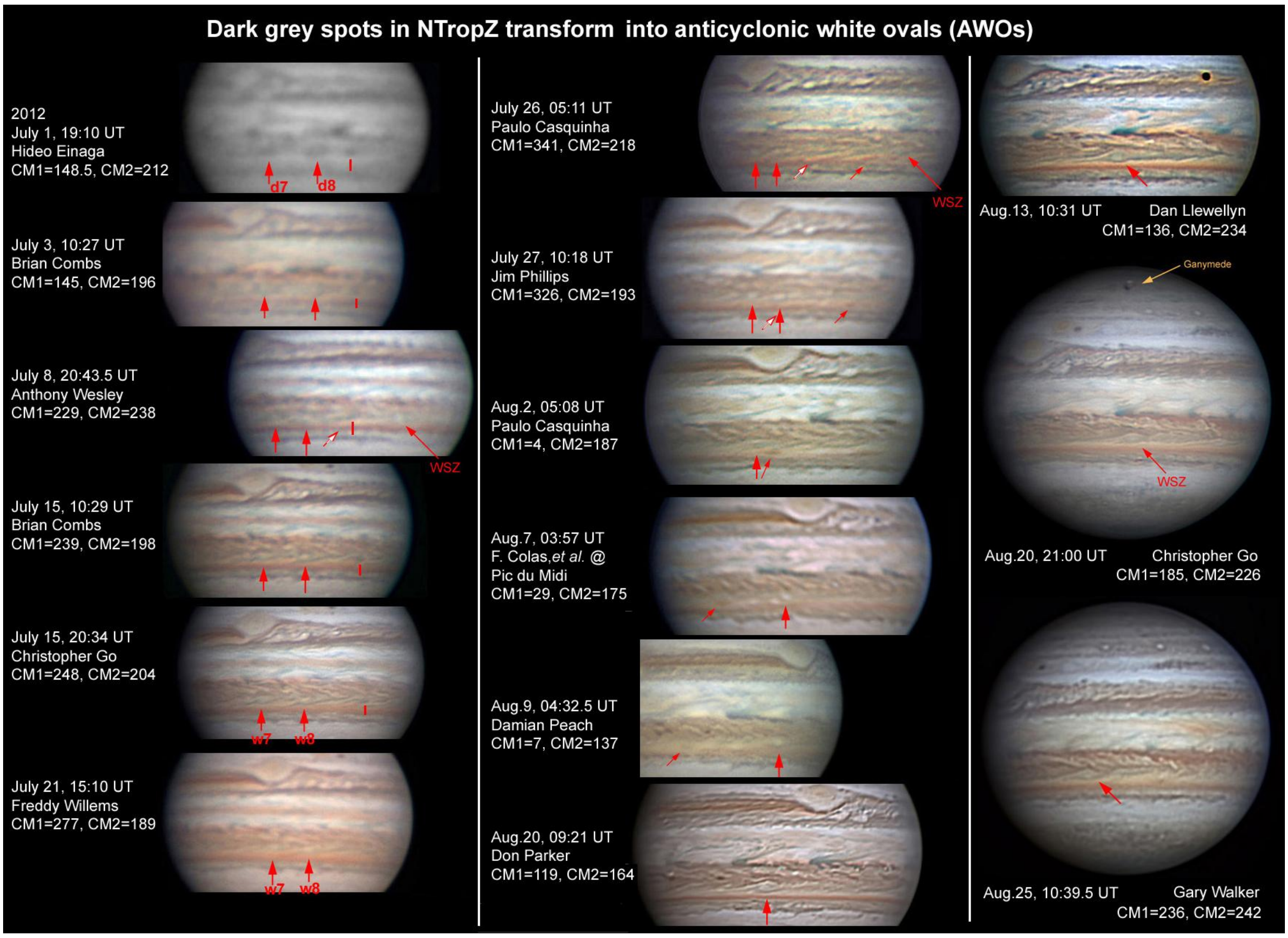

**Figure 8.** Images showing how dark grey patches eventually transform into AWOs WSA & WSZ. d7 and d8 are two large dark grey patches on July 1; by July 3, each has been largely replaced by a bright white spot (w7, w8) with dark grey stuff around it which is reduced to grey streaks by July 15. After this, the two white spots converge, and probably merge, as there is a single AWO (WSA) in August. Vertical red line marks another dark grey spot on July 1 which dissipates as a long grey streak. WSZ was a dark spot on July 3 (not shown), but has a bright outburst on July 8. Thereafter it is grey again in July, but has a light core in August. Adapted from larger sets in Ref.3f.

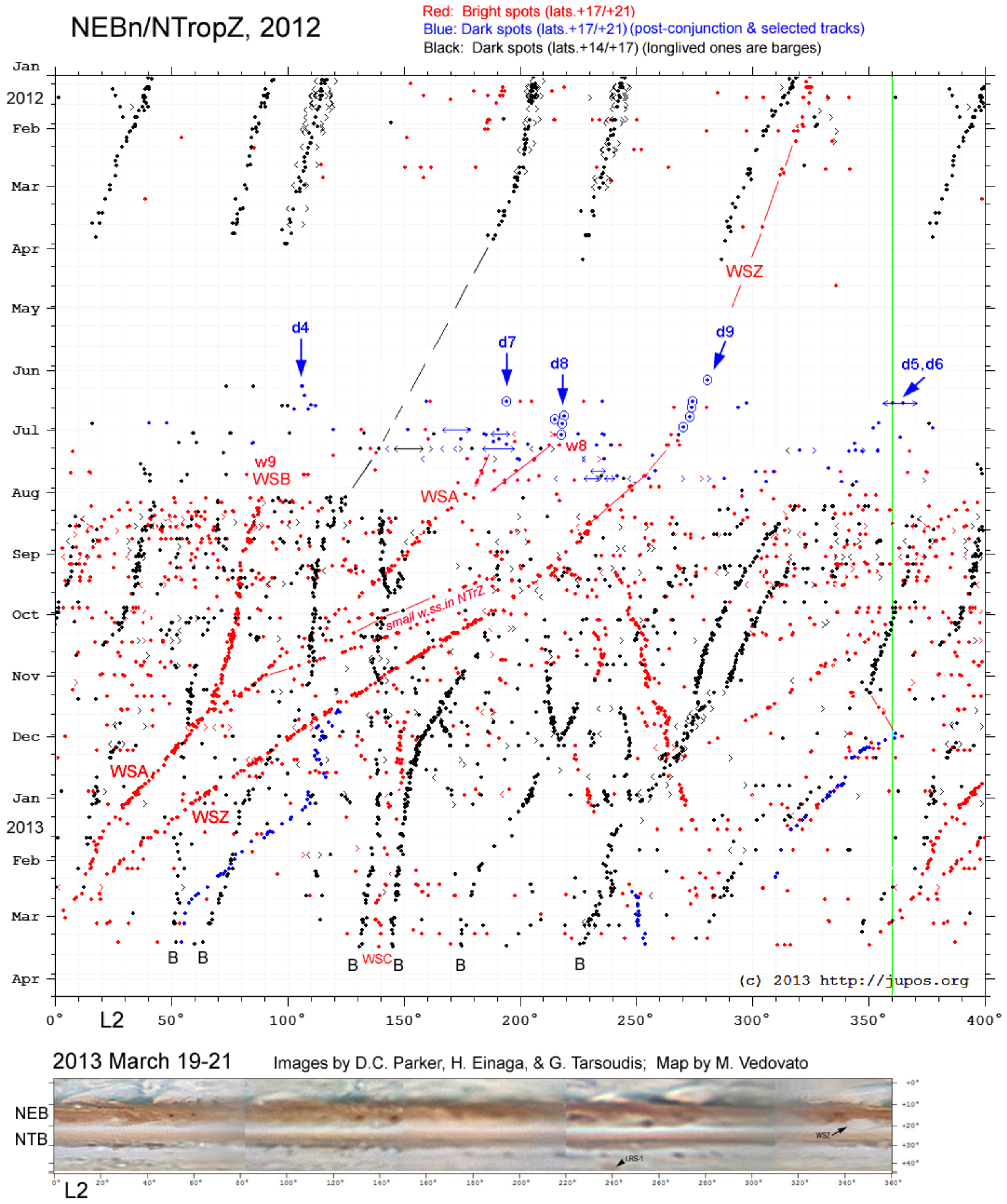

**Figure 9.** JUPOS chart of longitude vs time for NEBn/NTropZ features, latitudes 14-21°N, in L2, updated to cover the whole apparition. B = barges. At bottom, map towards the end of the apparition, aligned with the chart.

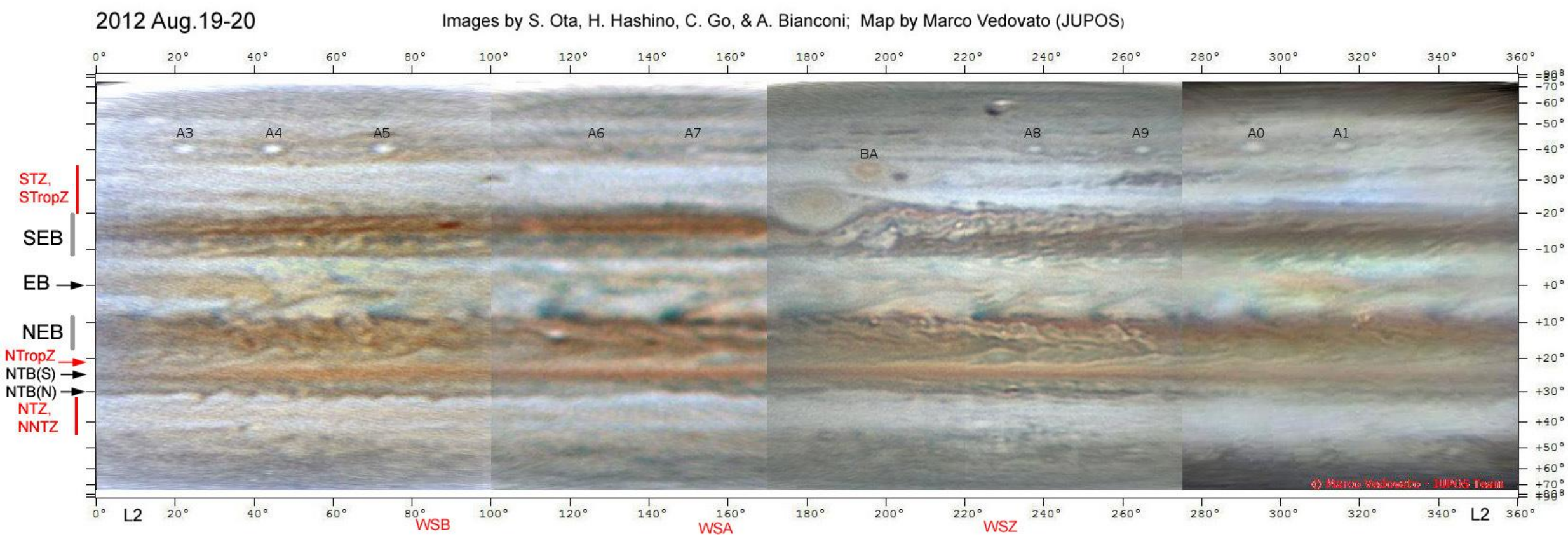

**Figure 10.** Map on 2012 Aug.19-20.

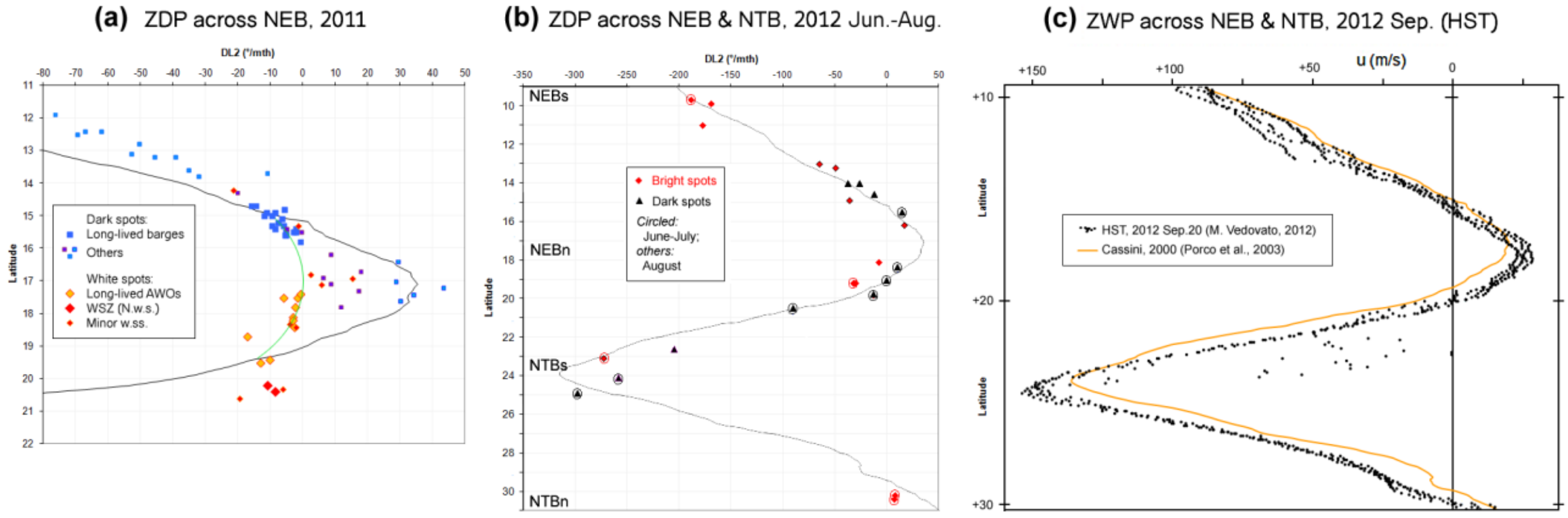

**Figure 11.** Zonal drift profiles (ZDP, for individual spots) and zonal wind profiles (ZWP, for cloud textures) before and after the upheaval, plotting drift or wind speed vs planetographic latitude. In each case the continuous curve is the ZWP from Cassini [ref.44], for comparison.

(a) ZDP in the 2011/12 apparition, before the outbreaks [from Ref.3a]. Points for WSZ are unusually far north as they represent a small white spot in its northern part. Green curve is a fit to the ZDP for major barges and ovals.

(b) ZDP in 2012 June-August [from Ref.3f]. Circled points are from June-July, others from August.

(c) ZWP from Hubble images on 2012 Sep.20, made using WinJUPOS by Marco Vedovato [from Ref.45].

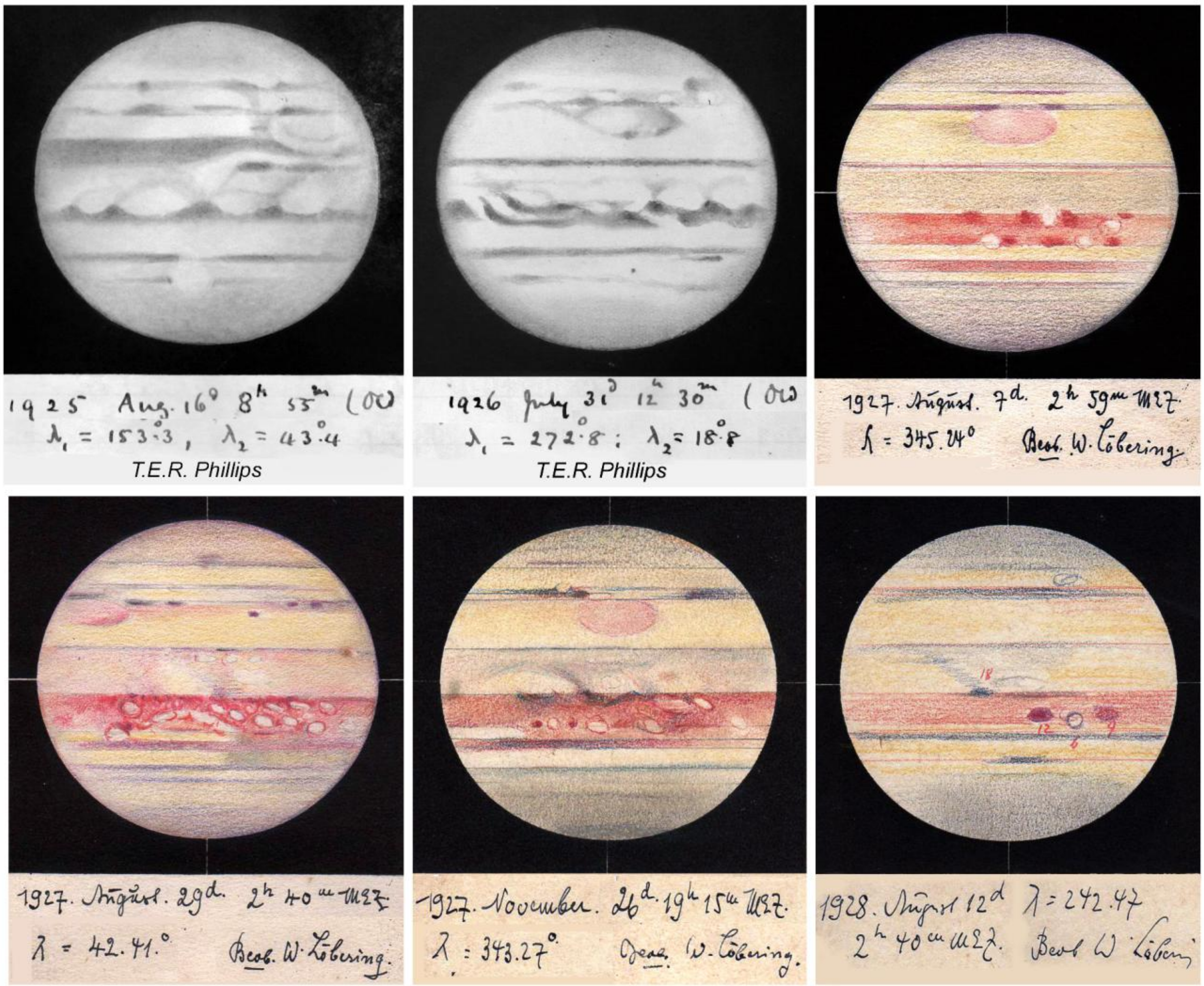

**Figure 12.** Original drawings in 1925-1928, from series by T.E.R. Phillips (monochrome, from BAA archives) and W. Löbering (colour, by courtesy of Frau Ulrike Toll and Hans-Jörg Mettig). Note the continuing expansion and disturbance of the NEB, and the reddish NTB in 1927.